\documentclass[11pt]{article}
\usepackage{Blois,epsfig,psfig}

\def\Journal#1#2#3#4{{#1} {\bf #2}, #3 (#4)}

\def\PLB{{\em Phys. Lett.}  B}
\def\PRL{\em Phys. Rev. Lett.}
\def\PRD{{\em Phys. Rev.} D}
\def\ZPC{{\em Z. Phys.} C}
\def\EPJC{{\em Eur. Phys. J.} C}
\def\MPLA{{\em Mod. Phys. Lett.} A}

\def\be{\begin{equation}}
\def\ee{\end{equation}}
\def\bea{\begin{eqnarray}}
\def\eea{\end{eqnarray}}

\begin{document}
\vspace*{2cm}
\begin{center}
\Large{\textbf{XIth International Conference on\\ Elastic and Diffractive Scattering\\ Ch\^{a}teau de Blois, France, May 15 - 20, 2005}}
\end{center}

\vspace*{2cm}
\title{ON THE ENERGY DEPENDENCE OF DIS IN THE DIFFRACTION REGION OF LOW X}

\author{ D. SCHILDKNECHT }

\address{Fakult{\"a}t f{\"u}r Physik, Universit{\"a}t Bielefeld, 
Universit{\"a}tsstrasse 25\\
D-33615 Bielefeld, Germany}

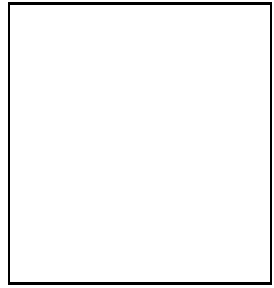
\maketitle\abstracts{
The energy dependence of the virtual photoabsorption cross section in
deep inelastic scattering (DIS) at low $x = Q^2/W^2 < 0.1$ may be described
in terms of the saturation scale that in our approach depends on the
energy, $\Lambda^2_{sat} = \Lambda^2_{sat} (W^2)$. We briefly summarize our
recent findings that allow us to predict the exponent $C_2$ of 
$\Lambda^2_{sat} (W^2) \sim (W^2)^{C_2}$ in agreement with the previous
result obtained from fitting the experimental data. The exponent $C_2$ 
depending on the relative magnitude of the longitudinal and transverse
contribution to the structure function, direct measurements of the longitudinal
part are urgently needed.}

At low $x \cong Q^2/W^2 < 0.1$, in deep inelastic scattering (DIS), the 
photon fluctuates  into an on-shell quark-antiquark 
vector state \cite{Sakurai,Gribov} that
interacts via two gluons \cite{Low} with the proton in the virtual-photon
forward-scattering Compton amplitude. The structure of the two-gluon-exchange
amplitude \cite{Nikolaev} implies a dependence of the virtual photoabsorption
cross section on the effective value of the gluon transverse momentum,
$\vec l_\bot$, thus introducing a dependence on a novel scale, relevant
at low $x << 1$, into the scattering process. In our representation 
\cite{Surrow,Diff2000,Ku-Schi} of the virtual photoabsorption cross section
and the structure function $F_2 (x, Q^2)$, the ``saturation scale'',
$\Lambda^2_{sat} (W^2)$, depends on the energy $W$,
\be
\Lambda^2_{sat} (W^2) = \frac{1}{6} \langle \vec l^{~2}_\bot \rangle =
\frac{1}{6} B^\prime \left( \frac{W^2}{1 GeV^2} \right)^{C_2},
\ee
and the fit to the DIS data gave \cite{Surrow}
\begin{eqnarray}
C^{exp}_2 & = & 0.27 \pm 0.01 \nonumber \\
B^\prime  & = & 0.340 \pm 0.63 GeV^2.
\end{eqnarray}
The photoabsorption cross section depends on the single scaling variable
\cite{Diff2000,Surrow}
\be
\eta = \frac{Q^2 + m^2_0}{\Lambda^2_{sat} (W^2)}.
\ee
A fundamental question concerns the magnitude of $C_2$ that is
responsible for the drastic rise of the structure functions $F_2 (x, Q^2)$
with decreasing $x$ at fixed values of $Q^2$ observed \cite{ZEUS} at
HERA.

We recently found \cite{Kuroda} that a representation of our approach
based on the interaction of a $q \bar q$ color-dipole vector state in
terms of the dual language of sea-quark and gluon distributions allows
one to determine the value of the exponent $C_2$. In terms of the
sea-quark, $x \sum (x,Q^2)$, and the gluon distribution, $xg (x,Q^2)$,
our approach contains the proportionality
\begin{eqnarray}
x \Sigma (x,Q^2) & = & \frac{4}{3 \pi} (2r + 1) \alpha_s (Q^2) x 
g (x, Q^2) =  \nonumber \\
& = & \frac{(2r+1)}{6 \pi^3} \sigma^{(\infty)} \Lambda^2_{sat} (W^2).
\end{eqnarray}
The constant $r$ is related to the ratio of the transverse-to-longitudinal
photoabsorption cross sections, at large $Q^2 >> \Lambda^2_{sat} (W^2)$,
\be
r = \frac{\sigma_{\gamma^*_T} (W^2, Q^2)}{2 \sigma_{\gamma^*_L} (W^2, Q^2)},
\ee
and, consistent \cite{Schi} with the available experimental information, 
$r = 1$ was assumed in the (successful) fit to the experimental 
data yielding (2).

The evolution of the sea-quark distribution, and consequently of 
$F_2 (x, Q^2)$, at low $x << 1$ and sufficiently large $Q^2$ is in
good approximation determined by the gluon distribution alone \cite{Prytz},
\be
\frac{\partial F_2 \left( \frac{x}{2}, Q^2 \right)}{\partial \ln Q^2} =
\frac{R_{e^+e^-}}{9 \pi} \alpha_s (Q^2) x g (x,Q^2).
\ee
Expressing $F_2 (x, Q^2)$ in terms of the sea-quark distribution,
\be
F_2 (x, Q^2) = \frac{R_{e^+e^-}}{12} x \Sigma (x, Q^2),
\ee
where $R_{e^+e^-} = 3 \sum_f Q^2_f = 10/3$, and making use of the 
proportionality (4), the evolution equation (6) becomes a constraint on
the $W^2$ dependence of the saturation scale,
\be
(2r+1) \frac{\partial}{\partial \ln W^2} \Lambda^2_{sat} (2W^2) = 
\Lambda^2_{sat} (W^2).
\ee
With the power law (1), one finds 
\be
(2r + 1) 2^{C_2} C_2 = 1,
\ee
or
$$
C_2^{theor.} = \cases{
0 & $(r >> 1)$, \cr
0.276 & $(r = 1)$, \cr
0.65 & $(r = 0)$.\cr
}
\eqno (10)
$$
The theoretical result (10) is in surprisingly good agreement with the
result (2) of the fit to the experimental data, based on $r = 1$. Moreover,
according to (10), the exponent determing the growth of $F_2 (x,Q^2)$ at
fixed $Q^2$ with decreasing $x$ is strongly correlated with the magnitude
of the longitudinal photoabsorption cross section relative to the 
transverse one. An approximately vanishing longitudinal  part of the
photoabsorption cross section, $ r >> 1$, only allows for a very weak
growth of the cross section and of $F_2 (x, Q^2)$ with decreasing $x$ at
fixed $Q^2$. In contrast, a dominating longitudinal cross section,
$r = 0$, implies an extremely strong growth of $F_2 (x, Q^2)$ with
decreasing $x$ at fixed $Q^2$.

Measurements allowing for the separation of the transverse and the 
longitudinal part of $F_2 (x, Q^2)$ are urgently needed to further
illuminate this interesting correlation between their relative magnitude
and the growth of $F_2 (x, Q^2)$ with decreasing $x$.


\section*{Acknowledgments}
This work was supported by Deutsche Forschungsgemeinschaft under
grant Schi 189/6-1.

\end{document}